%% file: Big_Data_Engineering.tex
\documentclass[11pt, twoside]{article}
\pdfoutput=1
\usepackage{amsmath}
\usepackage{amsthm}
\usepackage{booktabs}
\usepackage{caption}
\usepackage{enumitem}
\usepackage{makecell}
\usepackage{subfigure}
\usepackage{array}
\usepackage{arydshln}
\usepackage[dvipsnames]{xcolor}
\usepackage{multirow}
\usepackage{pifont}
\usepackage[margin=1.1in]{geometry}
\usepackage{tabto}
\usepackage{array}
\usepackage{setspace}
\usepackage{listings}
\usepackage{array}
\usepackage{tikz}
\usetikzlibrary{shapes, arrows, calc, positioning}
\usepackage{calc}
\usepackage{algorithm}
\usepackage{algpseudocode}
 \usepackage{color,soul}
\usepackage[colorlinks=true, citecolor=blue]{hyperref}
\usepackage{cleveref}
\usepackage{txfonts}

\theoremstyle{definition}

\makeatletter
\newcommand\footnoteref[1]{\protected@xdef\@thefnmark{\ref{#1}}\@footnotemark}
\makeatother

\hypersetup{%
    pdftitle={Database (Lecture) Streams on the Cloud --- An Experience Report on Teaching an Undergrad Database Lecture during a Pandemic},%
    pdfauthor={Jens Dittrich, Marcel Maltry},     
    pdfsubject={Databases},   
    pdfcreator={Jens Dittrich, Marcel Maltry},   
    pdfproducer={\LaTeX}, 
    pdfkeywords={Database, online teaching, streaming lectures, youtube},%
    pdfdisplaydoctitle=true,%
    pdfstartview=Fit,%
    colorlinks=false%
}

\begin{document}
\sloppy

\title{Database (Lecture) Streams on the Cloud\\---\\ An Experience Report on Teaching an Undergrad Database Lecture during a Pandemic}

\newcounter{savecntr}
\newcounter{restorecntr}
\author{Jens Dittrich\qquad
Marcel Maltry \\
Big Data Analytics Group\\
Saarland University\\
Saarland Informatics Campus\\
\url{https://bigdata.uni-saarland.de}
}

\date{\today}

\maketitle

\begin{abstract}
This is an experience report on teaching the undergrad lecture `Big Data Engineering' at Saarland University in summer term 2020 online. We describe our teaching philosophy, the tools used, what worked and what did not work. As we received extremely positive feedback from the students, in the future, we will continue to use the same teaching model for other lectures.
\end{abstract}

\input{body}

\end{document}

%% file: body.tex
\section{Background}
\label{intro}

In February 2020, many German universities decided to move to virtual teaching due to the unfolding Covid-19 pandemic and the public lockdown. As the summer term approached, lecturers had to decide relatively quickly how to `give all lectures online'. This included our CS department and our research group at Saarland University (Big Data Analytics Group). We already had quite some experience with teaching online:
In 2013, we started creating and using screencast videos for our lectures. From the beginning, we made our videos publicly available on YouTube.
Our channel\footnote{\url{https://youtube.com/jensdit}} currently has 11,000 subscribers and is one of the most successful database channels worldwide.
We already reported on that endeavor and in particular our inverted classroom philosophy in a previous report~\cite{DBLP:journals/dbsk/Dittrich14} in this journal.
However, since then, and given the experience with numerous inverted lectures, we decided to revisit all design decisions taken in the past. 

In this experience report, we describe what we did and why we did it. For each decision, we briefly discuss the alternatives and their pros and cons.
We do not claim that our model is the single right way to teach online.
There is neither a single right way to teach well offline nor online; there are a zillion different ways.  The challenge is to find the teaching style that fits you best. In summer 2020, we found a teaching style that fits well for us --- and as we learned from the students' evaluations --- it also fits for our students. With this work we would like to inspire others seeking to find the right bits and pieces for their own teaching.

This paper is structured as follows: We explain our teaching philosophy in Section~\ref{sec:teaching}. Our teaching material is discussed in Section~\ref{sec:material}. The hardware and software used are presented in Section~\ref{sec:technology}. In Section~\ref{sec:tutorials}, we proceed to discuss tutorials. We summarize advantages and issues with our teaching model in Section~\ref{sec:prosandcons}. Finally, we give an outlook to future plans in Section~\ref{sec:futureplans}.

\section{Teaching Philosophy}
\label{sec:teaching}

A major and common pitfall in online teaching is to \textit{first} start discussing technical tools, and specific software, and \textit{only then} consider the pedagogy and teaching philosophy behind it as a second thought. Like that, technology and software will define the pedagogy.

So let's first think about the teaching philosophy and pedagogy. There are two extremes: 

\begin{enumerate}
\item \textbf{Flipped (aka blended or inverted) teaching.} In flipped teaching, students receive material for self-study ahead of time.
The actual class is then used as a lab to work on exercises with the professor and Ph.D.~students\footnote{See our tutorial video \url{https://youtu.be/RRqoQAaeGCc} (in German) on how to organize a flipped classroom successfully.}.
\item \textbf{Live Teaching.} The other option is to run the lecture live over the Internet. This may be done as an online version of the physical lecture where content is `streamed' from the lecturer at the speaker's desk to the people in the lecture hall.

\end{enumerate}
Of course, all kinds of hybrids in-between these two extremes are possible. However, a major concern with all these approaches is the degree of interaction with and among the students.
A considerable strength of flipped teaching is the lab which is heavily interactive and perceived to be very useful by many students.
However, during the self-paced self-study phase, questions to the lecturer are not possible: the students are basically alone with the material. If they are stuck, issues might not be resolved quickly.

In contrast, a major strength of live teaching is the potential interactiveness: the lecturer can spontaneously ask questions, form short working groups, resolve urgent issues immediately, react to witty comments, and then depart in completely different directions, and so on. Like that, live teaching really becomes a live event that goes beyond a unidirectional stream of content from a lecturer to an audience.

So what is the `right' solution here? Again, there is no single `right' solution here. We debated back and forth and eventually decided to go with live teaching. Besides the potential of interactive elements, one reason for this decision was also the perceived high effort for creating video material for an entire lecture in a short period of time.

\section{Teaching Material}
\label{sec:material}

Until 2013, our lecture relied on material and examples from~\cite{KemperEickler15}. In 2013, we created our own material based around a hypothetical photo agency.
In 2019, we again created new material from scratch based on the following idea: we did not want to primarily organize the lecture along different topics anymore (first ER, then relational model, then relational algebra, etc.) as typically done, as we felt that this did not necessarily motivate students to fully appreciate the power of database technology.

\begin{figure}
\begin{center}
\includegraphics[trim = 0mm 0mm 0mm 0mm, clip, width=.63\textwidth,keepaspectratio,page=1]{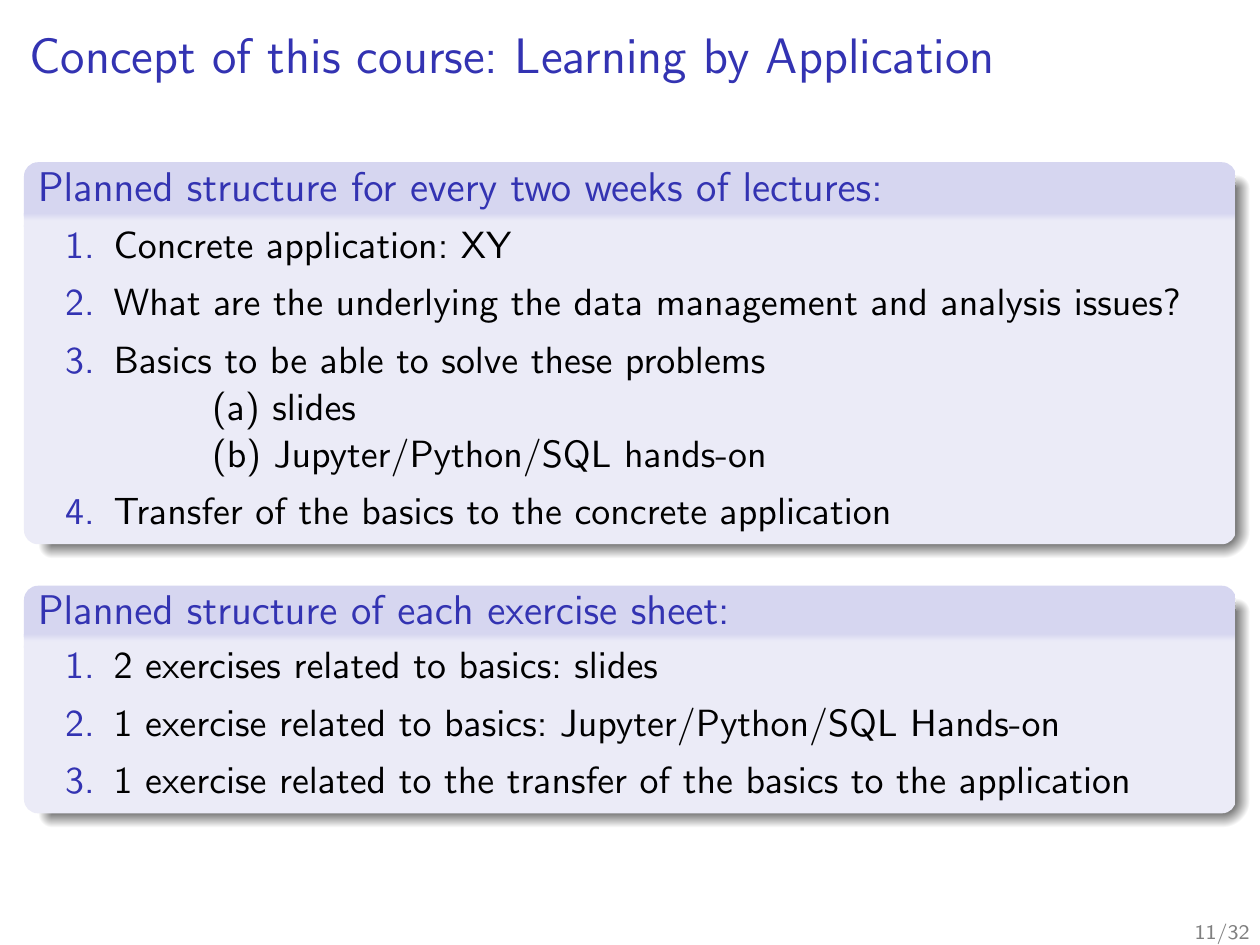}
\end{center}
\caption{Slide explaining the teaching philosophy of the course to students (translated)\label{fig:apporiented} }
\end{figure}

Instead, we decided to motivate certain topics by different applications. Every two weeks, we would pick a different application. This means, we would first show an application or problem and then showed how certain techniques from the database world would solve exactly that problem. Then, we would spent quite some time to make the transfer: how does that technology help in this application? This principle structure is summarized in~Figure~\ref{fig:apporiented}.

Table~\ref{tab:apps} shows the applications discussed and their mapping to topics. 

\begin{table}
\begin{center}
\begin{tabular}{|p{5.6cm}|p{5.9cm}|}
\hline
\textbf{Topic} & \textbf{Learning objectives} \\ \hline\hline
Python (Part~1, videos and/or 5.5.) & basics, functions, functional programming \\\hline

IMDb (Part~1, 7.5.) &  data modeling, relational model \\\hline

Python (Part~2, videos and/or 12.5.) & object-orientation, unit tests and automatic testing \\\hline

 IMDb (Part~2, 14.5.)  &  relationale algebra\\\hline

 NSA (Part~1, 28.5.)&  introduction to SQL\\\hline

 NSA (Part~2, 4.6.) &  analytical SQL, big data arithmetics, big data vs privacy, counter-measures\\\hline

 Query optimization (Part~1, 18.6.) &automatic query optimization,  physical operators, heuristic optimization\\\hline

 Query optimization (Part~2, 25.6.) & cost-based optimization, join order, plan variants, pipelining, physical optimizations\\\hline

 Trade, banks, ticket system (Part~1, 2.7.)& database management system (DBMS),\newline transactions, serializability theory \\\hline

 Trade, banks, ticket system (Part~2, 9.7.)& two-phase locking (2PL), isolation-levels\\\hline

%

 Summary  (16.7.)& \\\hline
\end{tabular}
\end{center}
\caption{Course agenda and their learning objectives. In 2019, we used one additional application: data journalism.
There we explained graph databases and security issues like SQL injection.\label{tab:apps}
}
\end{table}

As Saarland University decided to shorten summer term 2020 by starting only in May rather than in April, we also had to shorten the material. In addition, all courses had to be designed to allow all students to start in May only. If we provided material before May, we had to make sure to go through it again in May.  Due to these additional constraints, we decided to offer material for students who did not know Python yet as we planned to use Jupyter notebooks to explain certain concepts in the actual lecture. That Python introductory material was then again repeated in two lectures in May.

Moreover, throughout the lecture we also recommended old videos from 2013/14 to students in case they wanted alternative explanations. For all material that we created, we paid attention that our notation was consistent with~\cite{KemperEickler15}  in order to enable students to easily lookup yet another explanation in that book.

Also note that we left out considerable material that we felt like is not up to the reality of modern databases anymore, e.g.~normal forms whose importance can be debated in the light of modern non-scalar SQL-types\footnote{See this talk by Markus Winand \url{https://youtu.be/swR33jIhW8Q} for a great explanation of this.}.

All videos, including the older ones from 2013/2014 are publicly available on our YouTube-channel\footnote{\url{https://youtube.com/jensdit}}.

The pdfs of our slides are available through our website\footnote{\url{http://datenbankenlernen.de}}. If you are a lecturer and want to have access to the sources, send us an email.

We also created Jupyter notebooks\footnote{\url{https://github.com/BigDataAnalyticsGroup/bigdataengineering}} which were shown in actual lecture and used for the exercises.

\section{Technology}
\label{sec:technology}

In this section, we describe the hardware and software we used for our lecture.

\subsection{Hardware}

In terms of hardware, we experimented a lot until converging on the following setup. We used an existing 2016 MacbookPro, a 32 inch monitor, a thunderbolt dock (Elgato Thunderbolt 3 Pro Dock), a dynamic microphone with internal pop-filter (Rode Procaster), a mike preamp (TritonAudio FetHead), a sound interface (Focusrite Scarlett Solo 3rd Gen), and two speakers (Yamaha HS7 Standard). That's it! An existing iPad pro was not used in the lecture\footnote{An existing APS-C camera (Sony alpha 6000 with 1.8/24 Zeiss lens) powered by a fake battery adapter (SONY AC-PW20), was connected via an HDMI video card (Pengo hdmi USB-C 3.0 video capture-card), but so far not really used for the lectures, only for video meetings.}. The office used for streaming was a standard office room without any extra soundproofing (one exception: see Section~\ref{sec:OBS}). 

With this setups the sound quality of the audio is awesome and makes a huge difference over any setup we used before. In particular, previous issues with background noise are gone. The same holds for the camera setup which outperforms any dedicated webcam we tried.

\subsection{Internet Connection}
An important and surprising issue we ran into early was how to connect to the Internet. Initially we had quite some issues when connecting the computer via WiFi.
After experimenting with several setups, it became clear that WiFi is simply not stable enough for live streaming purposes.
This is not so much a problem of bandwidth but of channel conflicts and electromagnetic interferences from other devices, as well as latency issues.
Though we carefully debugged the WiFi connection, however, eventually we concluded 
that a wired ethernet connection should be preferred over WiFi whenever possible.
The same holds for audio and video conferences.

As LAN was not available in some home offices, we used a LAN via power option  (AVM FritzPowerline 510E Set).
Though powerline may lower the available bandwidth, it is much more stable than WiFi.
In general, in terms of Internet bandwidth note that even a relatively weak but stable uplink of 2Mb/sec may be enough to stream a lecture that only shows slides in the video stream. If you want to stream camera input, you need a higher bandwidth though.

\begin{figure}
\begin{center}
\includegraphics[trim = 0mm 0mm 0mm 0mm, clip, width=.63\textwidth,keepaspectratio,page=1]{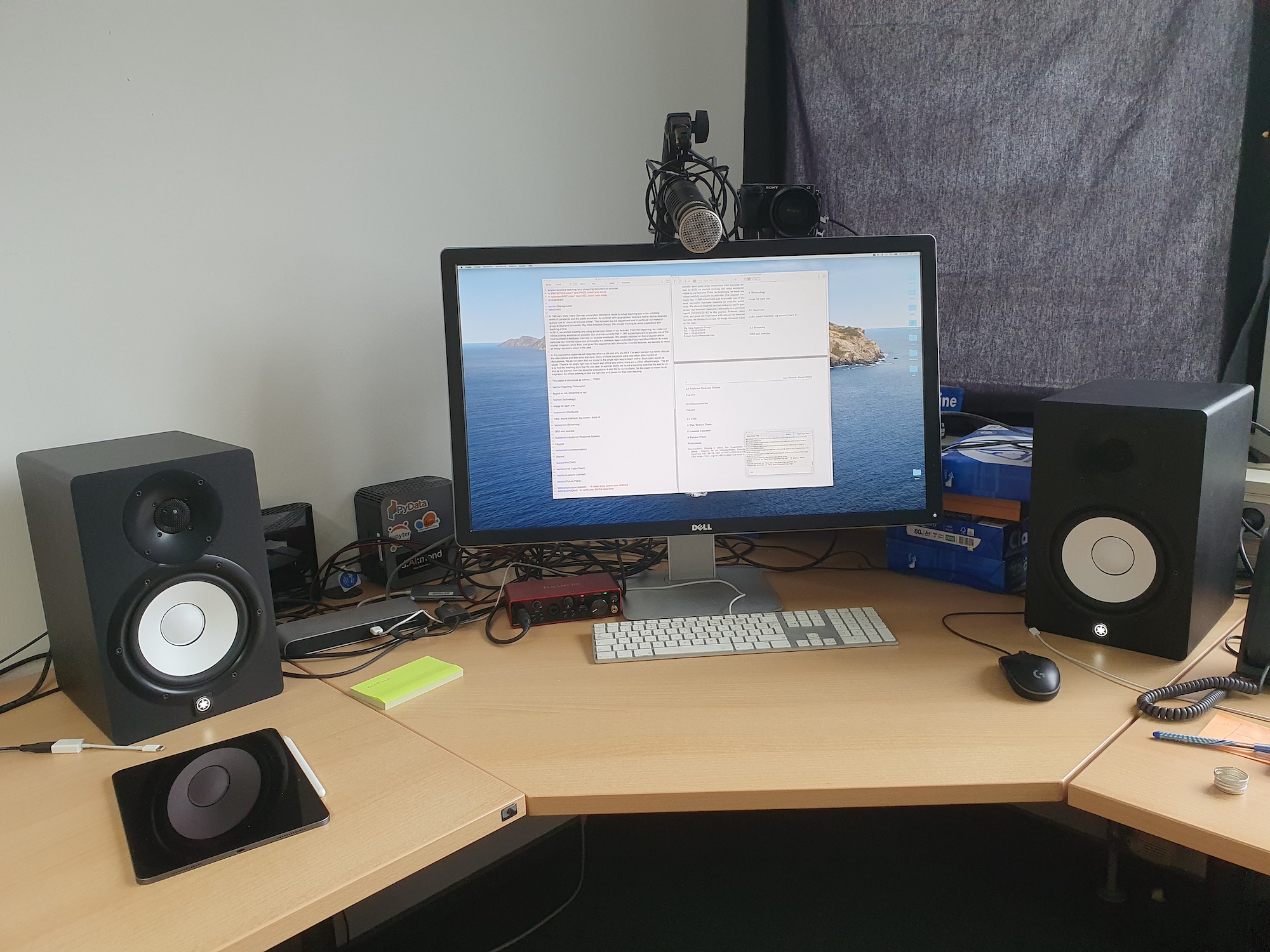}
\end{center}
\caption{Hardware setup (Macbook Pro not shown)}
\end{figure}

\subsection{Zoom, Teams or YouTube Streaming?}
\label{sec:OBS}

Initially we planned to use \textbf{Zoom} for the actual lecture and additionally stream the Zoom lecture to YouTube. This can directly be done from any Zoom meeting. However, shortly before the semester started we decided against using Zoom for several reasons: (1)~ongoing discussions on privacy and data protection issues with Zoom, (2)~relatively poor audio/video quality due to heavy lossy compression, (3) impossible to make the YouTube stream publicly available if at any time clear text names or webcam videos from students can be seen in the stream.

Similar concerns are true for \textbf{MS Teams}. After some brief investigation it turned out that the privacy and security issues with MS Teams were at least comparable to Zoom or even worse, e.g.~SSO with cleartext university password on microsoft.com website.
In addition, we experienced that the video and audio quality of MS Teams is simply not acceptable: we observed super-heavy compression artifacts, dichotomy of picture and sound. In addition, we found the UI of MS Teams to be extremely confusing.

\begin{figure}
\begin{center}
\includegraphics[trim = 0mm 0mm 0mm 0mm, clip, width=.63\textwidth,keepaspectratio,page=1]{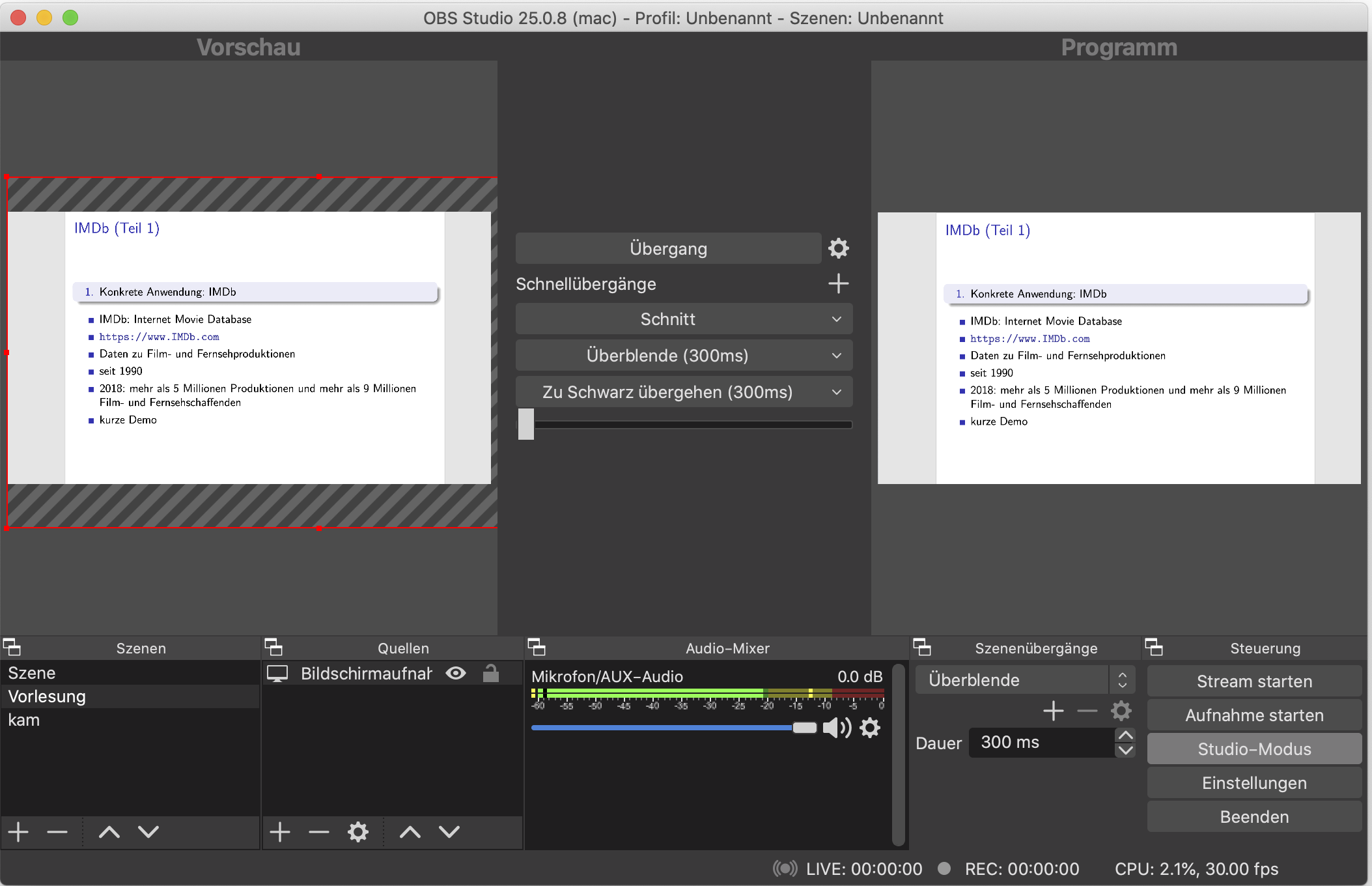}
\end{center}
\caption{OBS: Open Broadcaster Software was used to stream directly to YouTube\label{fig:OBS}}
\end{figure}

For the lecture, we decided to use a solution that fixes all of the above issues: \textbf{YouTube}'s livestream feature. In order to send a stream from your computer you have to install a streaming software locally. There aren't many options.
We decided to use \textbf{Open Broadcaster Software (OBS)}\footnote{\url{https://obsproject.com/}}, see Figure~\ref{fig:OBS}. Commercial options are also around, e.g.~Wirecast. OBS worked well for our purposes.
The only problem we ran into was that OBS is not really optimized for macOS. This results in a relatively high CPU load which in turn leads to notable fan noise.
We had to fix this by positioning the Macbook as far away from the microphone as possible and improvising some soundproofing between the Macbook and the microphone (a chair covered by a thick blanket).

By default, the head of the YouTube stream is cached to be able to cope with network failures. We observed a \textbf{latency} of about 10 seconds. We configured this to a shorter value to allow for more interaction with students (see Section~\ref{sec:ars})\footnote{We observed a lower lower bound for the stream latency of 2.5 seconds.}.

We also disabled the \textbf{livestream chat} to avoid spam; \textbf{video comments} were allowed though.

This setup has many advantage: we received great audio and video quality (1080p).
We could stream audio and video without any notable compression artifacts or distorted audio.
As the stream is automatically archived while being streamed, students can also pause the video, watch it time-shifted or at any point in time later.
In addition, all video format issues for different devices are automatically handled by YouTube.
Moreover, we do not put any extra load on the university network (if a video is watched from outside the network) which would have to be paid by our university (if billed per volume which is true for Saarland University).

\subsection{Audience Response System}
\label{sec:ars}

To allow for interaction with the students, we required an audience response system. We wanted something simple: a moderated chat where students could type in questions, a moderator (the tutor in chief) would select suitable questions, and forward them to the lecturer who would then drag the question to the stream and answer them live during the lecture.

\begin{figure}
\begin{center}
\includegraphics[trim = 0mm 0mm 0mm 0mm, clip, width=.68\textwidth,keepaspectratio,page=1]{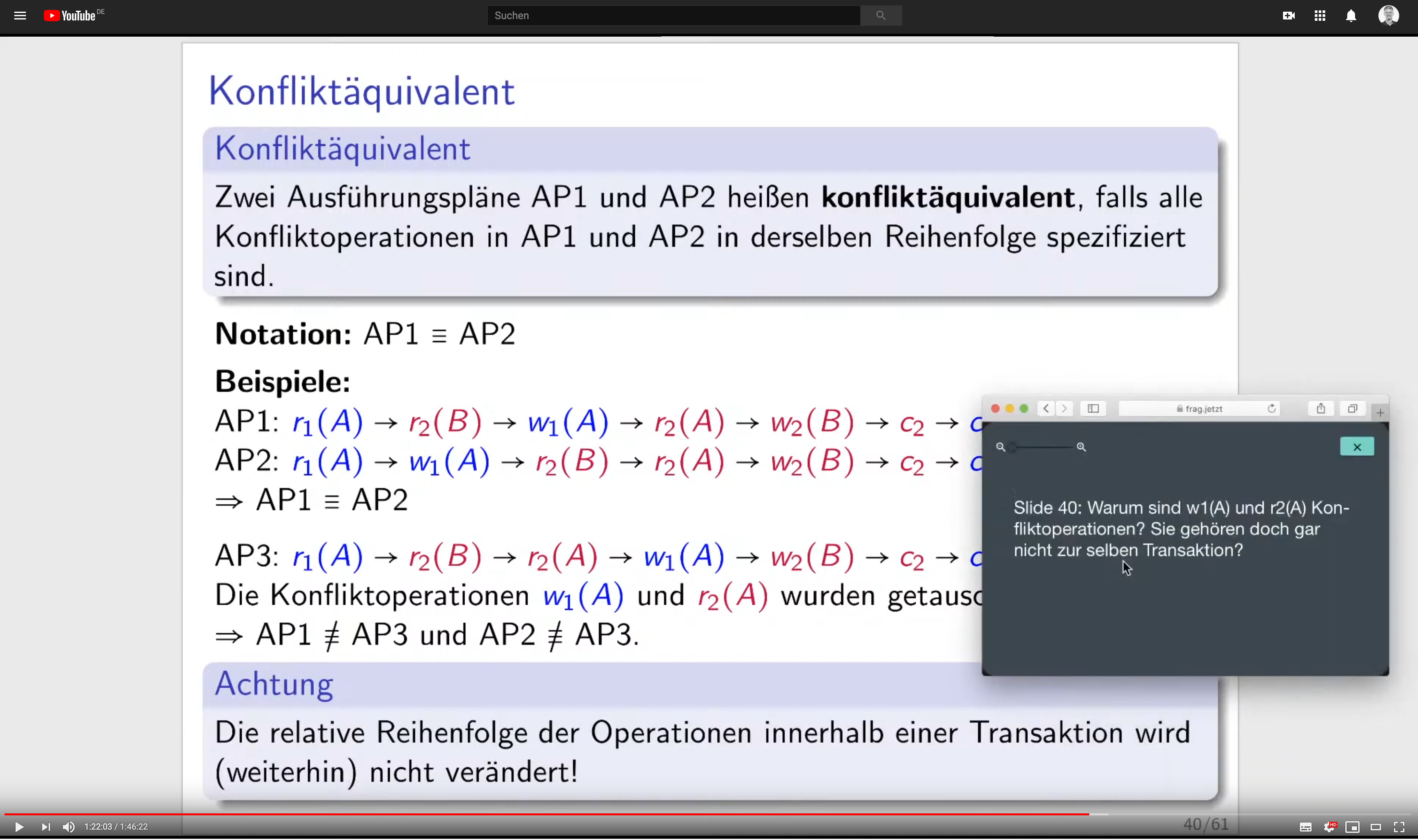}
\end{center}
\caption{Dragging a question from the audience asked in frag.jetzt into the livestream\label{fig:fragjetzt}}
\end{figure}

Many different tools are available in this space, and after some investigation we decided to use frag.jetzt\footnote{\url{http://frag.jetzt}}. It worked well for us. For this tool only the lecturer has to log in to make sure that for each lecture students can use the same URL.
See Figure~\ref{fig:fragjetzt} for an example.

The only problem we encountered with this tool is the missing possibility for the lecturer to ask back within the tool.

\subsection{Content Management System (CMS)}

As lecture system, i.e.~the system where students register, where all links to all materials are available, and where homework is handed in, we used an existing internal system.
This system has been in use at the department for more than ten years for physical courses as well. It is in many dimensions superior to other systems like Moodle (which we used for three years).
The biggest advantage is simplicity. In contrast, Moodle's UI is too complex and confusing.
It is sometimes hard to find things and in the student's evaluations it was often heavily criticized.

\subsection{Communication Tools}

Throughout the course, we used additional tools when needed.
In the following, we summarize three tools and provide use-cases for each of them.

We used \textbf{Discord} for office hours, tutorials, and mini office hours during the lecture break and right after each lecture. Discord is an audio conference tool widely used in gaming community.
The idea is to define virtual rooms which depending on your access rights, users are allowed to join anytime.
Once you enter a room, anyone in that room can hear what you say and you can hear immediately what others in that room are saying.
Though, at first, this sounds a bit like a standard video conference tool, in fact, it feels more like walking through different (physical) rooms.
Discord also allows for screen-sharing and video camera support. Discord was popular among students due to its familiarity and popularity in the gaming community. To join our server, students needed an invitation link that was only visible for students registered in our CMS. 

We `\textit{boosted}'\footnote{This is a fancy marketing term used by Discord to describe that we pay a monthly fee to get better services.} our Discord server to improve screen-sharing capabilities with a better resolution.
In addition, we installed a paid third-party Discord bot called VoiceMaster+.
It allows students to create their own temporary voice channels. 
This was  used by students to watch the lecture stream  `together'  to have a bit of a lecture hall feeling, in tutorials to create breakout sub-rooms, and for students' learning groups. 
Moreover, we also implemented multiple-choice questions for text channels (Pollbot). This was heavily used in the tutorials.

We also considered using Zoom or MS Teams for tutorials (cf.~Section~\ref{sec:OBS}).
We announced initially that we would only consider these tools as a backup.
In the end there was no need for this and the tutors and most students were happy with Discord.
Sometimes we observed stability issues with Discord. In addition, screen sharing and video is still a bit buggy and far behind Zoom's abilities.  Still, Discord is a great and very helpful tool that will surely further improve in future versions.

We used our CMS allowing students to give \textbf{anonymous feedback}. This was used about ten times in total in the entire semester.

We provided a \textbf{textual forum}. For this we used Discourse\footnote{\url{https://www.discourse.org/}}.
We witnessed several interesting discussions and witty student's comments.
We made sure to answer questions quickly and the lecturer involved himself into answering the hard questions. 
Due to Covid-19, the forum was more frequently used than in previous years.

In the tutorials, our students also used a tool for \textbf{shared scribbling}: A Web Whiteboard (AWW)\footnote{\url{https://awwapp.com/}}.

Obviously, this sounds a bit like we used \textbf{a zoo of tools} --- which is actually true.
Even though this is only a very small zoo, it still is a zoo.
To fix that problem, we tried to integrate the tools wherever possible and make everything easily accessible.
In the end, all the student-facing tools were web-based anyways, so the divide among the different tools was not as severe as it would have been for different desktop applications.

\section{Tutorials, Assignments, and Office Hours}
\label{sec:tutorials}

In addition to the lecture, we wanted to give the students the opportunity to deepen their understanding of the course material and ask questions. Therefore, we offered tutorials, office hours, and assignments. In the following, we first discuss our requirements for each of the offers and then explain how we implemented these requirements with the tools discussed in Section~\ref{sec:technology}.

\subsection{Tutorials}

In the past, we offered tutorials for up to 30 students in seminar rooms at our university. Students could vote for their favorite time slots in our CMS and were then assigned accordingly. This year, tutorials were held on our Discord server with a similar number of participating students. For some time slots two tutorials were offered simultaneously.
Therefore, we created two categories, one for each virtual seminar room. Each category was structured as follows:
\begin{itemize}
	\item [] \textbf{Plenum voice/video channel}: Participants and tutor meet here at the beginning of the tutorial. The tutor shares screen, gives instructions, and moderates the discussion.
	\item [] \textbf{Temporary voice/video channels}: Students form groups of up to five to discuss exercises.
	\item [] \textbf{Plenum text channel}: Tutors can send files and give instructions in case students are not in the plenum voice channel. Students can ask the tutor to join their temporary channel for questions.
\end{itemize}

In previous iterations of this lecture, tutors presented the solutions to the assignment handed in most recently. However, we observed that students tended to be very passive and simply wrote down the solutions in these types of tutorials. In fact, we could have simply provided a prerecorded video explaining the solutions but we wanted the students to participate in the tutorial actively.

Therefore, we decided against presenting solutions and instead prepared additional smaller exercises that could be solved during the tutorials. These exercises were explained by the tutors in a presentation-style screencast. Depending on the type of exercise, they were either discussed in the plenum or students were given 10 minutes to form small breakout groups and solve the exercises. We prepared exercises of the following three types:
\begin{enumerate}
	\item \textbf{Multiple-choice exercises}: These exercises were usually discussed in the plenum. Tutors would use our Pollbot to post a poll on the text channel and students would vote by reacting to the post. These exercises were especially helpful to gain insight on which topics were already well understood.
	\item \textbf{Written exercises}: These exercises were solved in smaller groups using temporary channels. Students had to apply concepts from the lecture to small problems. They often used AWW to work together on a solution.
	\item \textbf{Discussion exercises}: We often asked open questions that engaged the students to participate in a discussion, either in the plenum or in small groups.
\end{enumerate}

We learned that providing exercises to be solved in smaller groups significantly helps students to actively participate in the tutorials. We plan to also keep this approach once tutorials are held physically at the university again.

\subsection{Assignments}

From previous iterations of the lecture, we learned that encouraging the students to continuously recap and apply the concepts presented in the lecture leads to successful participation in the course. Therefore, we made correctly solving the majority of weekly assignments a prerequisite for admission to the exam.

Assignments should not only verify that students understood the concepts but also challenge good students to think further. Thus, assignments were divided into three parts:
\begin{enumerate}
	\item \textbf{Application tasks}: Students had to directly apply concepts from the lecture to familiar problems.
	\item \textbf{Transfer task}: Students had to apply concepts to new problems and argue about implications.
	\item \textbf{Programming task}: Students had to implement algorithms in empty cells in the Jupyter Notebook from the lectures. We provided basic unit tests for them to check their implementations.
\end{enumerate}
Assignments were published on, handed in, and graded using our internal CMS after each hand-in.
Since we chose a different approach for our tutorials this year, all solutions to the assignments were also published on our CMS.

As the lecture is aimed at undergrad students, we wanted to support them in practicing teamwork. We decided that assignments could be handed in by groups of up to three members. Since meeting physically to work on the assignments was not an option, we created a workspace category on our Discord server where students could create temporary channels for video calls, sharing their screen, and discussing the assignments.

\subsection{Office Hours}

Two years ago, we introduced an office hour. In contrast to the tutorials that have a fixed structure, we wanted to provide the students with a more open offer to stop by, discuss individual problems, and get help with technical issues. We held office hours once a week for two hours in a seminar room. However, most of the time, participating students just wanted to solve the assignment on-site and have a contact person in case of difficulties.

Therefore, we decided to offer office hours of one hour each. Office hours were held by two tutors on our Discord server. Participating students would enter a waiting area (i.e.~a public voice channel) and tutors would then accompany them to a private table (i.e.~a temporary voice channel). The temporary channels were configured such that other students could not join them to avoid distractions and keep conversations private. If a student wanted to join the conversation, the tutor had to move the student actively to the channel. Students mainly used office hours to receive help with technical difficulties, mostly setting up our environment for the Jupyter Notebooks. They shared their screen in the temporary channel where tutors then guided them towards a solution. Having shorter virtual office hours also resulted in fewer students with concrete questions and problems taking part. 

\section{Advantages and Issues of the Model}
\label{sec:prosandcons}

In summary, we are very happy with our teaching model. We also received tremendous positive feedback from our students in two different course evaluations.

\subsection{Advantages}

\begin{enumerate}
\item We avoided many of the privacy issues with Zoom and Teams.

\item The streams and their recordings are easily accessible, support time-shifted viewing, and are automatically archived.

\item There is no need to organize video servers, no load on the university's network (if watched from outside uni), all audio/video-format issues are automatically handled by YouTube, e.g.~to watch videos on different devices.

\item As we added video agendas in the video's comment sections, students could easily re-find things in the archived streams.

\item The tutorials with Discord were sometimes perceived to be better(!) than physical tutorials. The main reason was that  when working in small teams, students were not distracted by noise from other groups.

\item The public videos on YouTube are a great advertisement for our department and university (and let's hope for relational database technology in general). 

\item Videos are replacing textbooks. Our YouTube videos are used by students from other universities: rather than learning material from classical textbooks, they now use videos to prepare for exams. 

\end{enumerate}

\subsection{Issues}

\begin{enumerate}
\item For the lecturer, teaching without seeing the students was extremely exhausting. The main reason is probably that important communication channels for feedback were missing: e.g.~the expressions on student's faces, grumbling, laughing, etc., see also ``When everyone laughs for themselves''~\cite{kuehl20}.
Teaching often felt like talking to a wall. Sometimes while teaching you forget that you are speaking to two-hundred people.

\item For students, teaming up was sometimes difficult if they were not physically present on campus. This is not so much a problem in the 4th semester (as in this lecture) if they already found peers to team up with. Yet, this seems to be a severe problem for 1st semester students who may not know many peer students yet.
\end{enumerate}

\section{Future Plans}
\label{sec:futureplans}

In future, though we are not planning to change much, we aim to further improve the concept and fix the remaining issues. Here are some thoughts.
\begin{enumerate}
\item How to fix the missing visual backchannel problem? It would help the lecturer a lot to be able to see at least the faces of a subset of the students \footnote{Did you ever see students in a physical lecture hall who completely covered their faces? This would not be acceptable in western societies, yet this is commonly accepted in the virtual world.}. One solution could be to offer some students to additionally join a video conference call (Zoom or Discord) where they \textit{must} switch their camera on. 
However, it is unclear what the incentive for students could be. In particular, that incentive should be something that does not penalize students not willing to join that video call. 

\item In online teaching, there is an epic debate on whether the lecturer should switch the video camera on. What is the added value of this? In this lecture, we did not use the camera. However, we believe that, in particular for new students, it may be more helpful than for older students. Therefore we are considering this for an upcoming first semester lecture.

\item Due to many requests on YouTube, in the long run, we want to make all material available in English.
\end{enumerate}

\bibliographystyle{spbasic}      
\bibliography{jens}   

%% file: Big_Data_Engineering.bbl
\begin{thebibliography}{3}
\providecommand{\natexlab}[1]{#1}
\providecommand{\url}[1]{{#1}}
\providecommand{\urlprefix}{URL }
\expandafter\ifx\csname urlstyle\endcsname\relax
  \providecommand{\doi}[1]{DOI~\discretionary{}{}{}#1}\else
  \providecommand{\doi}{DOI~\discretionary{}{}{}\begingroup
  \urlstyle{rm}\Url}\fi
\providecommand{\eprint}[2][]{\url{#2}}

\bibitem[{Dittrich(2014)}]{DBLP:journals/dbsk/Dittrich14}
Dittrich J (2014) {Die Umgedrehte Vorlesung - Chancen f{\"{u}}r die
  Informatiklehre}. Datenbank-Spektrum 14(1):69--76,
  \doi{10.1007/s13222-013-0143-9},
  \urlprefix\url{https://doi.org/10.1007/s13222-013-0143-9}

\bibitem[{Kemper and Eickler(2015)}]{KemperEickler15}
Kemper A, Eickler A (2015) Datenbanksysteme: Eine Einf\"{u}hrung, 10th edn. De
  Gruyter Studium, De Gruyter Oldenbourg

\bibitem[{Kühl(2020)}]{kuehl20}
Kühl S (2020) {Wenn jeder für sich allein lacht}. Forschung und Lehre (5),
  \urlprefix\url{https://www.forschung-und-lehre.de/lehre/wenn-jeder-fuer-sich-allein-lacht-2778/}

\end{thebibliography}
